\documentclass[prb,aps,twocolumn]{revtex4}
\usepackage{graphicx}
\begin{document}


\title{Mobility gap in fractional quantum Hall liquids: Effects of
disorder and layer thickness}

\author{Xin Wan$^{1,2}$}
\author{D. N. Sheng$^3$}
\author{E. H. Rezayi$^4$}
\author{Kun Yang$^5$}
\author{R. N. Bhatt$^6$}
\author{F. D. M. Haldane$^7$}

\affiliation{$^1$Institut f\"ur Nanotechnologie,
Forschungszentrum Karlsruhe, 76021 Karlsruhe, Germany}
\affiliation{$^2$Zhejiang Institute of Modern Physics,
Zhejiang University, Hangzhou 310027, P.R. China}
\affiliation{$^3$Department of Physics and Astronomy, California State University,
Northridge, California 91330, USA}
\affiliation{$^4$Physics Department California State University Los Angeles,
Los Angeles, California 90032, USA}
\affiliation{$^5$National High Magnetic Field Laboratory and Department of Physics,
Florida State University, Tallahasse, Florida 32306, USA}
\affiliation{$^6$Department of Electrical Engineering, Princeton University,
Princeton, New Jersey 08544, USA}
\affiliation{$^7$Department of Physics, Jadwin Hall, Princeton University,
Princeton, New Jersey 08544, USA}

\date{\today}

\begin{abstract}

We study the behavior of two-dimensional electron gas in the fractional
quantum Hall regime in the presence of finite layer thickness and
correlated disordered potential.  Generalizing the Chern number calculation
to many-body systems, we determine the mobility gaps of 
fractional quantum Hall states based on the distribution
of Chern numbers in a microscopic model. 
We find excellent agreement between experimentally measured activation
gaps and our calculated mobility gaps, 
when combining the effects of both disordered potential and layer thickness. 
We clarify the difference between mobility gap and spectral gap of 
fractional quantum Hall states and explain the disorder-driven 
collapse of the gap and the subsequent transitions 
from the fractional quantum Hall states to insulator.

\end{abstract}

\maketitle


\section{Introduction}

One of the most remarkable properties of two-dimensional electron gas
(2DEG) is the amazing precision of the Hall resistivity quantization
$\rho_{xy} = h/(ie^2)$ in a perpendicular high magnetic field at low
temperatures, regardless of materials, geometries, impurities, and
carrier concentrations of experimental systems. 
This phenomenon is known as the integer quantum Hall
effect~\cite{vonklitzing80} (IQHE) for integer $i$, or as the fractional
quantum Hall effect~\cite{tsui82} (FQHE) for certain fractional values
of $i$. 
At the Hall resistivity plateaus, the longitudinal resistivity
$\rho_{xx}$ vanishes at zero temperatures, 
but has an Arrehenius-type temperature $T$ dependence 
\begin{equation}
\rho_{xx} \propto \exp (-\Delta / 2 k_B T)
\end{equation}
where $k_B$ is the Boltzmann's
constant.~\cite{chang83,boebinger87,willett88}
The thermally activated behavior suggests that there is an activation
gap $\Delta/2$ in the excitation spectrum of each quantum Hall state. 
The gap has an origin of Landau level spacing in the
IQHE~\cite{prangebook} and a more
profound origin of electron-electron interaction in the
FQHE;~\cite{laughlin83} in fact, 
the existence of the activation gap leads directly to the Hall
resistivity quantization and the consequent incompressibility of the
corresponding Hall liquid at low temperature.

Theoretically, $\Delta$ in the FQHE is 
expected to be the creation
energy of a pair of free quasielectron and quasihole. In a pure system,
this is the asymptotic value of the excitation spectrum 
in the large momentum limit, found to be as large as 
$0.1 e^2/\epsilon l_B$,~\cite{haldane85,girvin85} 
where $\epsilon$ is the dielectric constant and $l_B = (\hbar c /
eB)^{1/2}$ the magnetic length. 
However, experiments~\cite{chang83,boebinger87,willett88} found much
smaller excitation gaps, presumably due to the reduction caused by 
the presence of disorder, the thickness of the 2DEG layer, 
and the mixing of Landau levels.
Yoshioka~\cite{yoshioka86} combined the effects of layer
thickness~\cite{zhang86} and Landau level mixing,~\cite{yoshioka84} 
and obtained reasonable agreement with experimental results in
high-mobility systems at large enough magnetic fields.~\cite{willett88} 

However, without taking into account the effects of disorder,
theoretical considerations cannot explain the vanishing 
activation energy below finite magnetic field (about 5
T).~\cite{boebinger87,willett88} 
Qualitatively, disorder broadens the quasielectron-quasihole excitation
band, leading to a reduction of the energy gap between the ground state
and excited states.~\cite{chang83}
MacDonald {\it et al.}~\cite{macdonald85} and Gold~\cite{gold86,gold87} 
considered the effects of disorder. 
Both theoretical approaches contain adjustable parameters, and more
importantly, fail to answer the following crucial questions: 
Which quasielectron-quasihole excitations are contributing to the
activated longitudinal resistivity?
What is the nature of the activation gap? 

To answer these questions, let us look at the single-particle
picture of the IQHE first. 
In a clean system with exactly $n$ (labeled from 0 to $n-1$) Landau
levels filled, the activation gap is obviously the Landau level spacing
$\hbar \omega_c$, where $\omega_c = eB/m^*c$ is the cyclotron
frequency. This involves the excitation of an electron in the $(n-1)$-th
Landau level to the $n$-th Landau level, or equivalently, the excitation
of a pair of electron in the $n$-th Landau level and hole in the
$(n-1)$-th Landau level. In the presense of disorder, each Landau level
is broadened into a Landau band (with bandwidth $2\Gamma$, say).
Due to Anderson localization, localized states exist in the tails
of each band, and delocalized states only exist in the center of the
band (where the mobility edge is). 
The excitation of an electron from a localized level to another
does not contribute to the longitudinal resistivity; only excitations
involving delocalized levels matter.  
Therefore, although the spectral gap in this case is reduced to $(\hbar
\omega_c - 2\Gamma)$, the energy to excite a pair of free electron
and free hole is still $\hbar \omega_c$ -
related to the mobility gap rather than the spectral gap of the system. 
In the FQHE, on the other hand, we have quasielectron and quasihole
excitations. Similar to electrons in the integer case, 
these quasiparticles can be trapped in their potential
valleys and become localized, 
thus do not contribute to the longtidinal resistivity. 
Therefore, we need to find, for the FQHE,
the mobility edge - the energy beyond 
which quasiparticle excitations are delocalized.

In the noninteracting IQHE, 
the calculation of topologically invariant Chern numbers has been 
established~\cite{thouless82,niu85,arovas88,huo92,sheng95,sheng96,yang97,bhatt02}
as a reliable way to obtain the Hall conductance, 
to measure the localization length critical exponent, and
to determine whether a single-particle state is localized or
conducting - thus where the mobility edge is.
Physically, the Chern number of a state is the (dimensionless) Hall
conductance, which can be derived from the Kubo formular, averaged over
boundary conditions of a finite system on a torus.~\cite{niu85} 
In addition, it has an elegant geometric interpretation as the integral 
of the curvature of the quantum state in the parameter space 
spanned by two angular parameters (twisted boundary conditions) - the first
Chern class of a U(1) principal fiber bundle on the
torus.~\cite{avron83,simon83,avron03} 
Chern numbers are topologically invariant 
under small perturbations of Hamiltonian, such as weak disorder, which 
allows the mobility gap to open for a finite range of magnetic field, 
leading to the plateau structure of the IQHE
as long as the Fermi level lies in the mobility gap.

However, the geometric interpretation of the Hall conductance leads to
an apparent controversy~\cite{niu85,avron85} in the FQHE. 
On the one hand, if a system exhibits the FQHE, the many-body ground
state of the system must be degenerate on a toroidal geometry; 
otherwise, gauge-invariance arguments 
produce only integral Hall conductance.~\cite{tao84} 
For a pure system at filling fraction $\nu = p/q$ ($p$ and $q$
relatively prime to each other), the
ground-state manifold is $q$-fold degenerate. Generically, these
$q$ degenerate states share a total Chern number $p$, regardless of the
Chern number carried by each state. 
Thus, the Hall conductance of the pure system is the average Chern
number of these $q$ states, fractionally quantized at $\sigma_H = pe^2/qh$. 
However, numerical calculations show that impurities lift the
degeneracy in a finite system,~\cite{zhang85} implying that the
Hall conductance would only be quantized to fractional values 
under superfluous conditions.~\cite{tao86}

Wen and Niu~\cite{wen90} proposed that states of a system exhibiting the
FQHE are topologically degenerate on a torus 
(in general, on high-genus Riemann surfaces) 
in the thermodynamic limit.
The ground-state degeneracy is, in fact, a signature of
the topological order of the bulk, invariant against weak but otherwise
arbitrary perturbations (including symmetry-breaking impurity
potential).~\cite{wenbook}
In a finite system, the quasidegeneracy of the states replaces the
exact degeneracy, which can be recovered in the thermodynamic limit. 
The topological degeneracy thus guarantees the Hall conductance to be
quantized at fractional values in the fractional regime even in dirty
systems, in which the topological Chern number for each many-body state
is well-defined. 
Based on these ideas, we performed
a numerical study of topological Chern numbers for 
the $\nu = 1/3$ FQHE in finite systems.~\cite{sheng03}  
The mobility gap for the FQHE can be determined from the distribution of
the Chern numbers of the quasidegenerate many-body states.
The results quantitatively explain the absence of the activation gap
of the FQHE due to disorder at small magnetic fields, as well as the
disorder-driven collapse of the gap and the subsequent
transition from the FQHE to insulator at higher fields, 
observed by various experiments.~\cite{chang83,boebinger87,willett88} 

In this paper, we further apply the method of Chern number calculation
to study the effects of layer thickness of 2DEG and correlated
potential on mobility gaps of FQHE systems. In Sec.~\ref{sec:model},
we introduce our microscopic model for 2DEG with disordered potential 
and finite layer thickness, and explain 
the Chern number calculation for FQHE systems.  
In Sec.~\ref{sec:thickness}, we discuss the
effects of layer thickness on the properties of 2DEG, in particular on
mobility gaps, which we compare to the activation gaps measured by
experiments. We then discuss the effects of correlated impurity
potential on mobility gaps in Sec.~\ref{sec:correlation} before we
summarize our results in Sec.~\ref{sec:summary}.

\section{Model and Method}
\label{sec:model}

We consider a two-dimensional (2-D) polarized interacting 
electron system on an $L_1\times L_2$ rectangular area 
with generalized periodic boundary conditions (PBCs)
\begin{equation}
T({\bf L}_j)\Phi({\bf r})=e^{i\theta_j}\Phi ({\bf r}),
\end{equation}
where $T({\bf L}_j)$ is the magnetic translation
operator and $j=1,2$, representing $x$ and $y$ directions,
respectively.
In the presence of a strong magnetic field, one can project 
the Hamiltonian of the system 
onto the partially-filled, lowest Landau level.
Therefore, we consider the following projected Hamiltonian 
in the presence of both Coulomb interaction
and disorder\cite{sheng03}
\begin{eqnarray}
H&=& {1 \over A} \sum _{i<j} \sum _{ \bf {q}}  e^{-q^2/2 }  V(q)
e^{i {\bf q} \cdot ({\bf R}_i -{\bf R}_j)} \nonumber \\
&+&\sum _{i} \sum _{ \bf q} e^{-q^2/4} U_{\bf q}
e^{i{\bf q }\cdot {\bf R}_i},
\end{eqnarray}
where ${\bf R}_i$ is the guiding center coordinate of the $i$-th
electron,
$U_{\bf q}$ is the impurity potential 
with wave vector ${\bf q}$, and 
\begin{equation}
\label{eqn:interaction}
V(q)={2\pi e^2 \over \epsilon q} F(q) 
\end{equation}
is the Fourier transform of the electron-elctron interaction. 
The factor $F(q)$ generalizes the Coulomb interaction to 
the case with finite electron layer thickness 
(to be explained in the next paragraph).
We use the Gaussian white-noise potential generated according to the
following correlation relation in $q$-space
\begin{equation}
\langle U_{\bf q}U_{{\bf q}'}\rangle=
{W^2 \over A} \delta_{{\bf q},-{\bf q'}},
\end{equation}
which corresponds, in real space, to 
\begin{equation}
\langle U({\bf r})U({\bf r'})\rangle
=W^2 \delta (\bf {r-r'}),
\end{equation}
where $W$ is the strength of the disorder
(in units of $e^2/\epsilon$)
and $A = 2 \pi N_s l_B^2$ is the area of the system. 
To study the effects of correlated potential, 
we also generate the Gaussian correlated random potential $U(r)$
according to 
\begin{equation}
\langle U_{\bf q}U_{{\bf q}'}\rangle = 
{W^2 \over A} \delta_{{\bf q},-{\bf q'}} e^{-q^2 \xi^2 / 2},
\end{equation}
which leads to
\begin{equation}
\langle U({\bf r}) U({\bf r}') \rangle 
= {W^2 \over 2 \pi \xi^2} e^{- |{\bf r} - {\bf r}'|^2 / 2 \xi^2},
\end{equation}
where $\xi$ is the characteristic correlation length. 
Note that in the
limit of $\xi \rightarrow 0$, we recover the Gaussian white-noise potential. 

To describe the thickness of a quasi-two-dimensional electron system, 
we employ the Fang-Howard variational wave function 
\begin{equation}
\phi(z) = (b^3/2)^{1/2} z e^{-bz/2},
\end{equation}
where $b$ depends on the material properties and the carrier density
of the system. 
The parameter $b^{-1}$ has the physical meaning of electron layer
thickness, thus we will use, unless otherwise specified, 
the dimensionless layer thickness $\beta = (bl_B)^{-1}$ hereafter.
The experimentally interesting range of parameter
is $\beta \sim 1$. 
For the Fang-Howard wave function,
the reduced Coulomb interaction in two dimensions becomes~\cite{zhang86} 
\begin{equation}
V(r)=(e^2 / \epsilon) \int_0^{\infty} dq F(q) J_0(qr),
\end{equation}
where 
\begin{equation}
F(q) = \left (1 + {9 \over 8} {q \over b} 
+ {3 \over 8} {q^2 \over b^2} \right ) \left(1 + {q \over b} \right )^{-3}
\end{equation}
and $J_0$ is the Bessel function of zeroth order. 
This factor $F(q)$, the same as in Eq.~(\ref{eqn:interaction}), 
softens the bare Coulomb interaction between electrons, 
especially at short distances. 

We diagonalize the Hamiltonian with Lanczos algorithm 
and compute the Hall conductance $\sigma_H$ through the Chern number
calculation, which offers an unambiguous criterion to distinguish
between insulating and current carrying states in an interacting
system.~\cite{arovas88,sheng03}
With a unitary transformation 
\begin{equation}
\Psi_k= \exp \left [-i \sum _{i=1}^{N_e} \left (\frac {\theta_1}{L_1} x_i 
+ \frac {\theta_2}{L_2}y_i \right ) \right ] \Phi_k,
\end{equation}
we can write the boundary-condition averaged Hall conductance
for the $k$-th many-body eigenstate $\Phi_k$ 
as $\sigma_H(k)= C(k)e^2/h$,
where the Chern number $C(k)$ for the state is 
\begin{eqnarray}
\label{integral}
C(k) ={i\over 4\pi} \oint_{\Gamma} d {\bf \theta} \cdot
\left [ \langle { \Psi_k |{\partial \Psi_k
 \over
\partial {\bf \theta}}\rangle -
\langle {\partial \Psi_k \over \partial {\bf \theta}}|
\Psi_k}
\rangle \right ].
\end{eqnarray}
Here, the closed path integral is carried out along the boundary
$\Gamma$ of the boundary condition space (the magnetic Brillouin zone)
$0 \leq \theta_1, \theta_2 \leq 2\pi$. 
$C(k)$ is exactly the Berry phase (in units of $2\pi$) accumulated for
the state when the boundary conditions evolve along $\Gamma$.
We separate the magnetic Brillouin zone into 
at least $25$ meshes depending on the system size and 
calculate the sum of the Berry phase from each mesh.
For the mesh sizes we choose, we find 
converged integer Chern numbers.
We emphasize that throughout the paper we use the rectangular geometry, 
which facilitates the calculation of Chern numbers. 
A heuristic but qualitative discussion 
on the ground-state splitting and localized quasiparticle excitations 
also exists for a spherical geomotry.~\cite{thouless89}

\section{Effects of Layer thickness}
\label{sec:thickness}

In this section, we discuss the effects due to the finite thickness of
the 2DEG for $\nu = 1/3$. 
We use the Fang-Howard variational wave function, introduced
in Sec.~\ref{sec:model}, to describe the electronic wave function in the
perpendicular direction. We consider the Gaussian white-noise
potential, and compare the disorder effects to the ideal two-dimensional
case.~\cite{sheng03} In the following subsections, we present results on
density-of-state, energy and energy split of ground states, spectral gap,
distribution of Chern numbers, and mobility gap. These results are
qualitatively similar for cases with and without finite layer
thickness.

\subsection{Density-of-states}

\begin{figure}
{\centering \includegraphics[width=7cm]{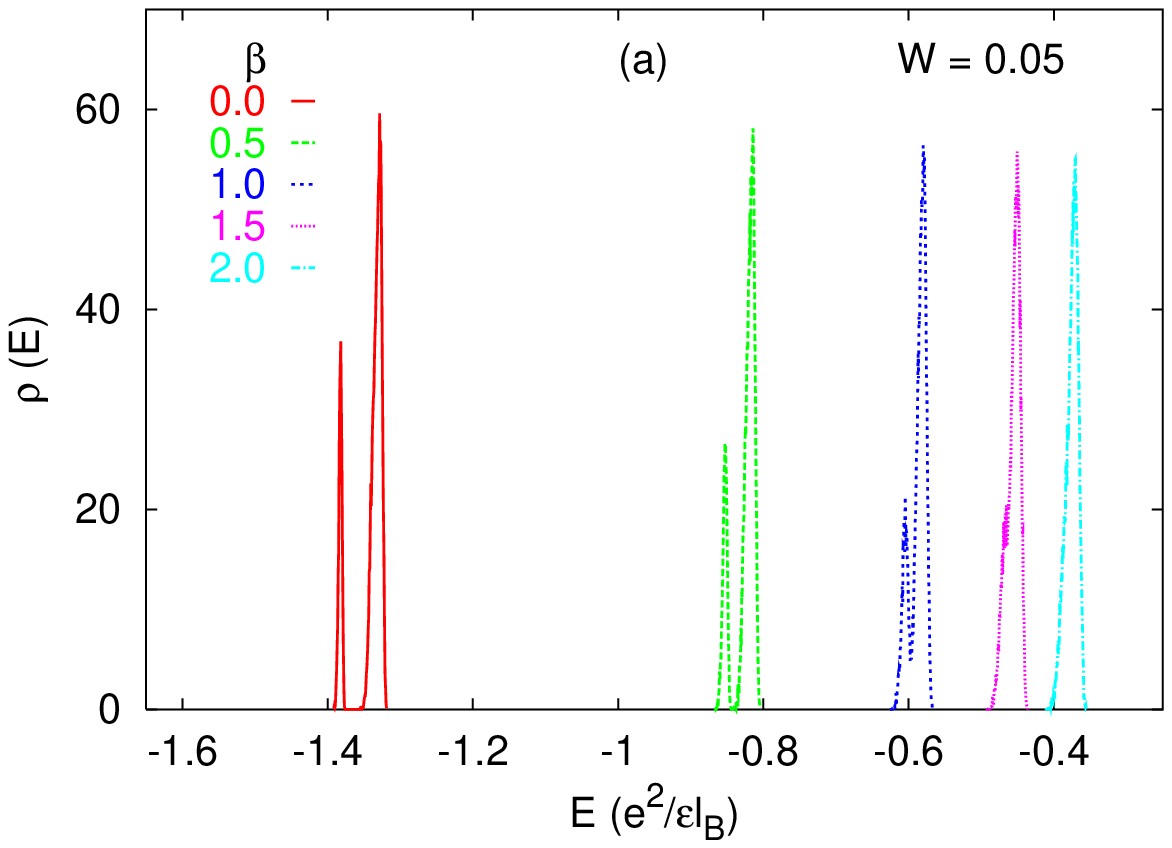} }
{\centering \includegraphics[width=7cm]{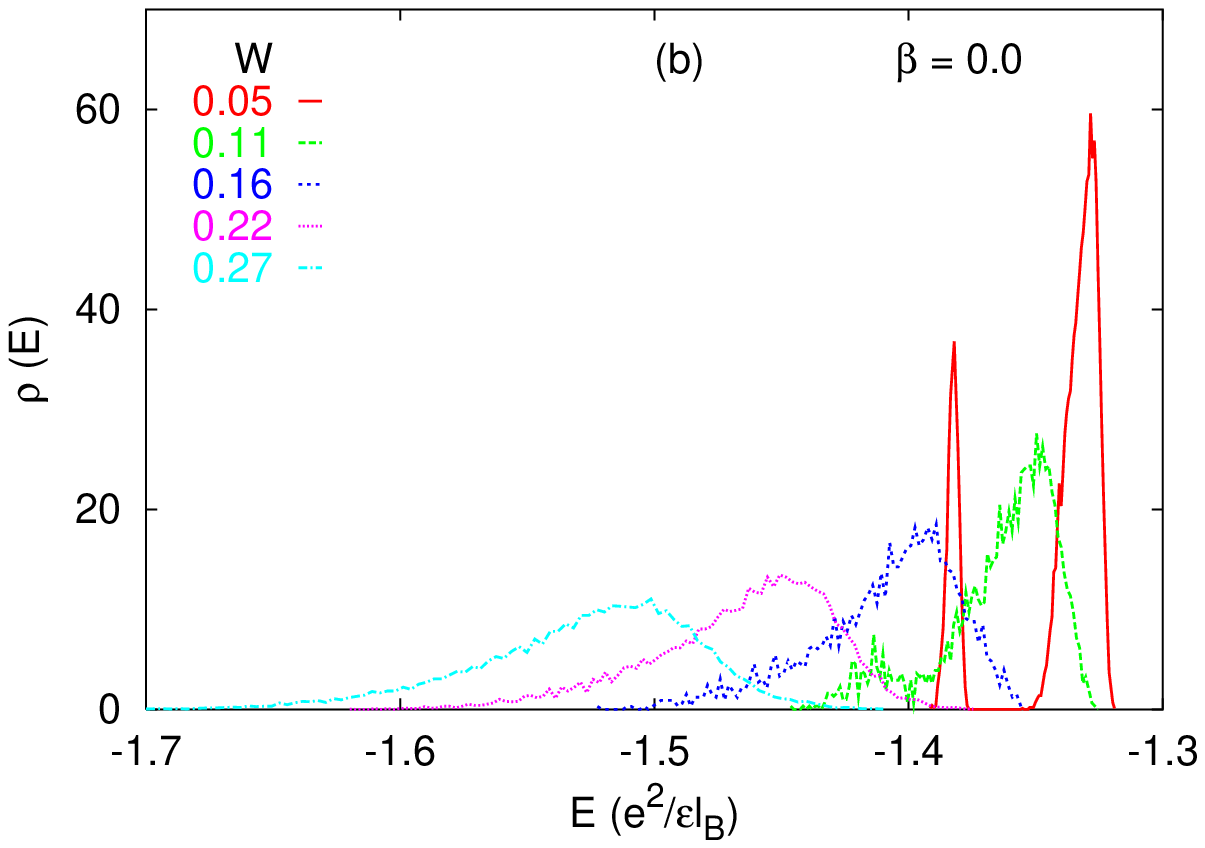} }
\caption{
\label{fig:dos}
Density of the lowest 15 many-body states for $N_e = 6$ at $\nu = 1/3$. 
(a) $W = 0.05$ with $\beta = 0.0$, 0.5, 1.0, 1.5, and 2.0. 
(b) $\beta = 0.0$ with $W = 0.05$, 0.11, 0.16, 0.22, and 0.27.  }
\end{figure}

We diagonalize the system to obtain up to 30 lowest states for each
sample for $N_e = 3$-8. 
Figure~\ref{fig:dos}a shows the evolution
of the many-body density of the 15 states as the layer thickness changes in the
upper panel for weak disorder $W = 0.05$ and system size $N_e = 6$ . 
For $W = 0.05$ and $\beta = 0.0$, a spectral gap is visible
in the density-of-states with the low energy peak consisting of three
nearly degenerate states. 
As $\beta$ increases, the energies of states
becomes closer to zero while the spectral gap narrows, 
reflecting the softening of the Coulomb interaction. 
For $\beta = 2.0$, even a weak $W = 0.05$ is sufficient 
to destroy the spectral gap completely. For comparison, 
Fig.~\ref{fig:dos}b shows the evolution of the density of the 15 states 
as $W$ increases for zero layer thickness $\beta = 0.0$ 
in the lower panel. 
Disorder broadens the density of states and the low energy peak merges
with other states at $W > 0.11$ for $N_e = 6$. 

\subsection{Ground-State Energy}

The effects of the finite layer thickness on the ground-state energy of the
fractional quantum Hall effect has been considered by MacDonald and
Aers.~\cite{macdonald84} 
In the absense of disorder, they found that the the Laughlin state
energy at $\nu = 1/3$ for $\beta = 1.0$ reduces to 0.576 
of its value for $\beta = 0.0$.
Chakraborty~\cite{chakraborty86} found a similar reduction ratio of
0.579 using the hypernetted-chain method. 

\begin{figure}
{\centering \includegraphics[width=7cm]{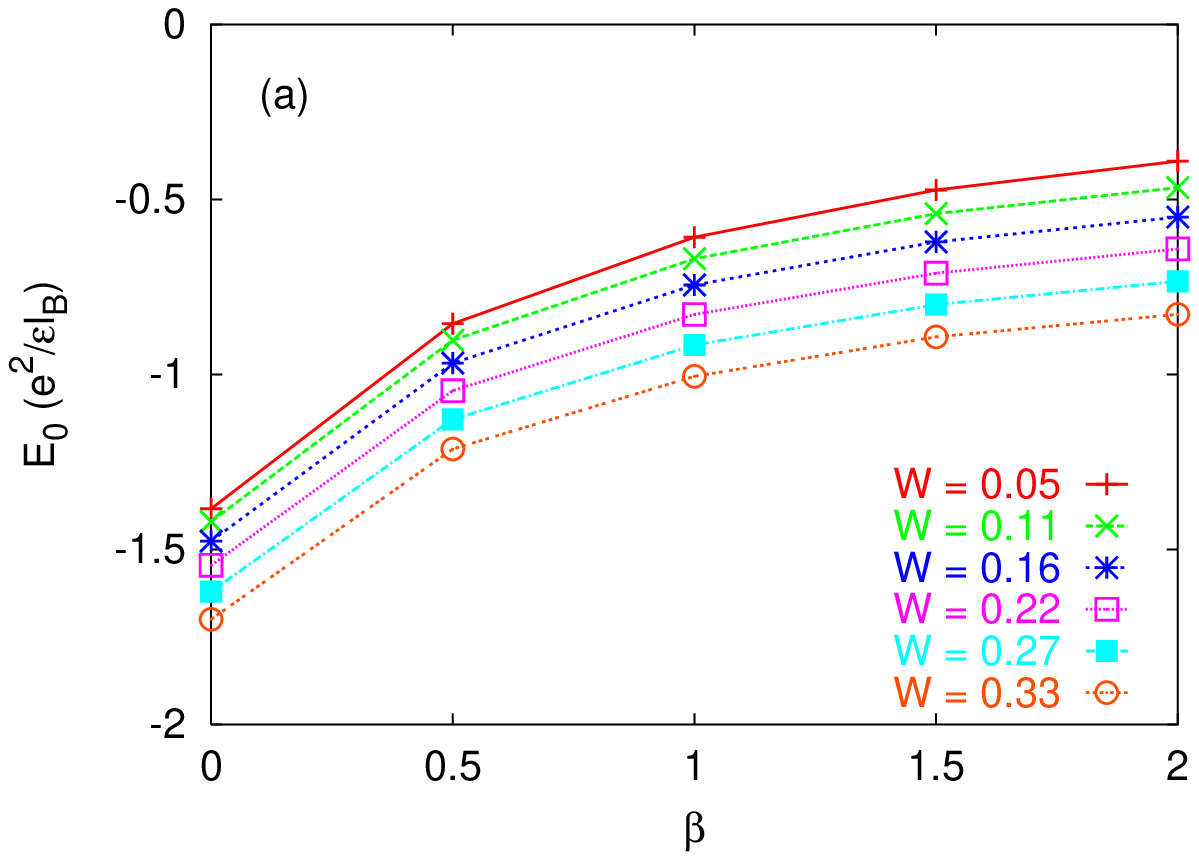} }
{\centering \includegraphics[width=7cm]{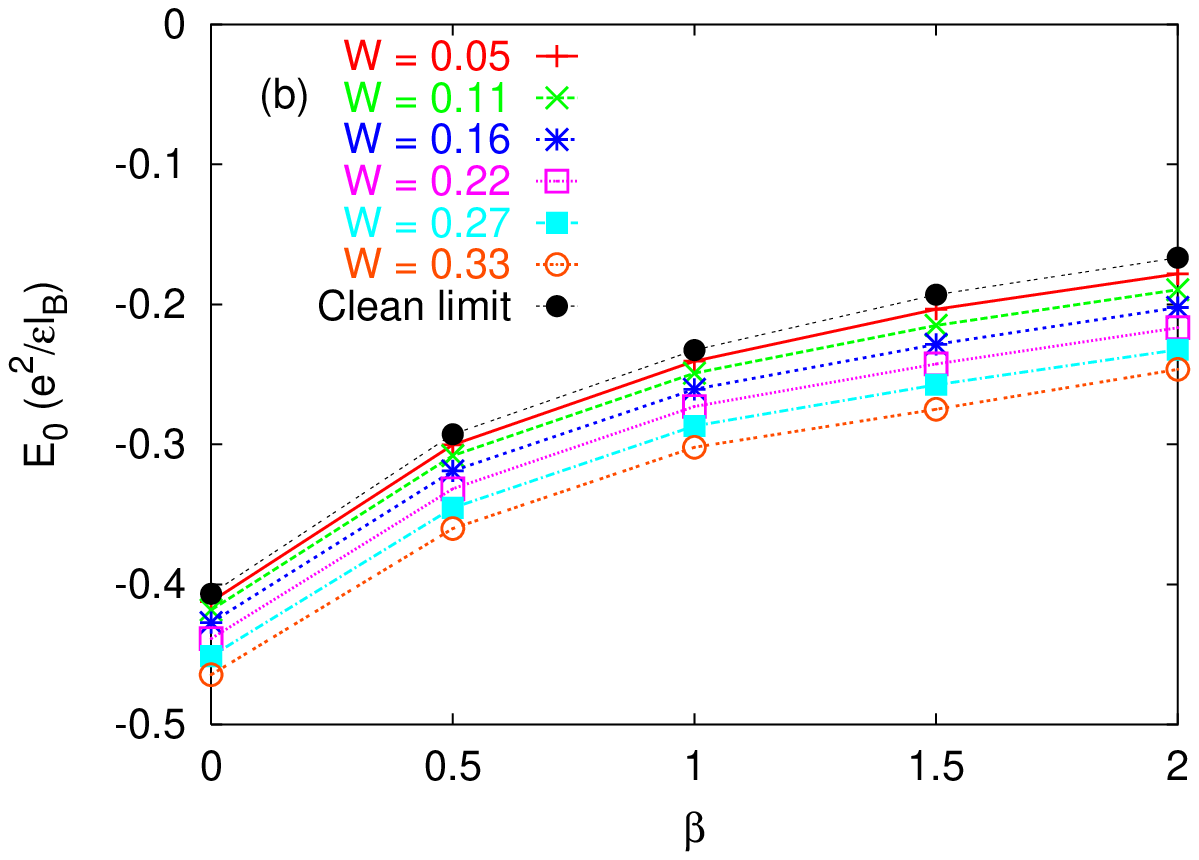} }
\caption{
\label{fig:gse}
(a) Ground-state energy (excluding interaction between electrons and
their images) as a function of $\beta$ for $N_e = 6$ at $\nu =
1/3$. (b) Ground-state energy per electron,
extrapolated to the thermodynamic limit, as a function of
$\beta$ for $\nu = 1/3$. 
}
\end{figure}

In the presence of disorder, we find a similar reduction in magnitude 
of the many-body ground-state energy $E_0$ in finite systems. 
Figure~\ref{fig:gse}a shows the reduction of the ground-state 
energy as a function of $\beta$ for $N_e = 6$ at $\nu = 1/3$. 
Here, we defined the ground-state energy as the average energy of the
lowest three levels, which are topologically degenerate 
in the thermodynamic limit.
To compare our results with earlier
works~\cite{macdonald84,chakraborty86}, we first add, to the many-body
ground-state energy, a single-particle contribution from interactions of
an electron and its images due to periodic boundary conditions. 
This energy can be related to the Coulomb energy of the classical
square Wigner crystal with finite layer thickness.~\cite{fujiki92}
We then extrapolate our data for $N_e = 3-7$ electrons to the
thermodynamic limit according to 
\begin{equation}
E_0(N_e) = E_0(N_e \rightarrow \infty) + a_0 / N_e.
\end{equation}
The results for ground-state energy per electron are shown 
in Figure~\ref{fig:gse}b.
By extrapolating the results to the clean limit by a quadratic fit, we
obtain a reduction ratio of 0.572 for the ground-state energy from
$\beta = 0.0$ to $\beta = 1.0$.
The value is in good agreement with known results in the clean
case.~\cite{macdonald84,chakraborty86} 
In the presence of the impurity potential, disorder shifts the ground
energy state down, leading to a slightly larger ratio (e.g., 0.596 for
$W = 0.11$). 

In the current and earlier works~\cite{macdonald84,chakraborty86}, the ground
state energy is negative, due to the assumption of a uniform neutralizing
background charge, which cancels out the singular contribution of the
Coulomb interaction among electrons. Mathematically, we substract the
singular $q = 0$ contribution from the electron-electron interaction, as
illustrated in Appendix~\ref{sec:squarelattice}. 
Therefore, a reduction of $E_0$ in magnitude is, in fact, 
an increase of the ground
state energy, as layer thickness increases. 
This is mainly because the corresponding reduction of the
attractive interaction between electrons and the background charge 
negates the contribution from the softening of the electron-electron
repulsion. Meanwhile, $E_0$ decreases with increasing $W$ because
electrons take advantage of the negative potential region in the ground
state as shown in Fig.~\ref{fig:gse}. 

\subsection{Energy split of the ground states of finite systems} 

Wen and Niu~\cite{wen90} pointed out that the ground states of the
fractional quantum Hall state are degenerate on a torus even in the
presence of disorder. However, this is strictly valid only in the
thermodynamic limit. For a finite system on a torus of length $L$, the
ground states have an energy split, or a bandwidth $E_b$, of order 
\begin{equation}
\label{eqn:bandwidth}
E_b \sim e^{-L(m^* \Delta)^{1/2}},
\end{equation}
where $\Delta$ is the quasiparticle-quasihole pair creation energy and
$m^*$ is the effective mass of the quasiparticle. 
The energy split comes from the tunneling process that a
virtually created pair of quasiparticle and quasihole propogate in
opposite directions and annihilate on the other side of the torus. 

\begin{figure}
{\centering \includegraphics[width=7cm]{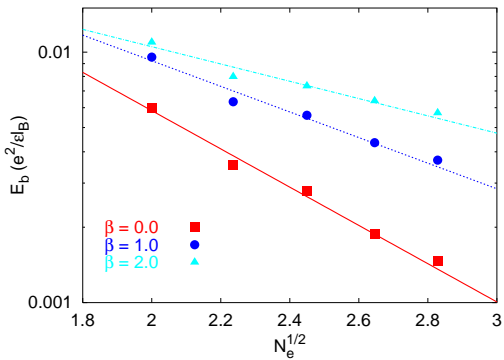} }
\caption{
\label{fig:bdw}
Energy split of ground states as a function of the square root of 
electron number (linear system size) on a semilog scale for various 
$\beta$ at $W = 0.05$ and $\nu = 1/3$. 
}
\end{figure}

In figure~\ref{fig:bdw}, we plot the bandwidth $E_b$ of the lowest three
states, which are well separated from all higher levels at $\nu =
1/3$ and $W = 0.05$, as a function of $\sqrt{N_e}$, 
proportional to the linear system size. 
For various $\beta$, the decrease of $E_b$ with $\sqrt{N_e} \sim L$ is
consistent with the exponential decrease in Eq.~(\ref{eqn:bandwidth})
expected by the theory,~\cite{wen90} 
as found in Ref.~\onlinecite{sheng03} for zero layer thickness. 
Though, we cannot completely rule out a power-law decrease 
of $E_b$ based on these finite-size values. 
We find $E_b$ increases with the layer thickness, again consistent with 
Eq.~(\ref{eqn:bandwidth}) assuming that the
quasiparticle-quasihole pair creation energy decreases with increasing
$\beta$, as the Coulomb repulsion between electrons reduces.
From $\beta = 0.0$ to 1.0, the slope in the semilog plot of
figure~\ref{fig:bdw} reduces by roughly 30\%. The reduction is, however,
smaller than that of either the spectral gap or the mobility gap which
we will discuss in the following sections. 

We also note that a weak periodic potential, instead of disorder, gives
rise to a similar exponential dependence of the splitting on system
size,~\cite{pfannkuche97} implying, again, a degenerate ground-state
manifold in the thermodynamic limit. 
The nontrivial ground-state degeneracy, existing on Riemann surfaces
with genus 1 (torus) or greater~\cite{wen90}, is a signature of 
topological order~\cite{wenbook} possessed by these states. 
The basic physics of the topological degeneracy and the finite-size 
splitting of the ground-state manifold in FQHE is the same as 
in chiral spin liquid, which is also a prototype for studying
topological order and quantum error-correcting code.~\cite{bonesteel00}

\subsection{Spectral Gap}

Since the ground-state manifold on a torus consists of three
quasidegenerate levels for $\nu = 1/3$, 
we define the spectral gap as the energy difference between the
third and the fourth lowest energy states
\begin{equation}
E_s = E(4) - E(3).
\end{equation}
At large disorder, $E_s$ simply becomes the energy level spacing, 
when the lowest three levels no longer form a separate band from higher
levels. 

\begin{figure}
{\centering \includegraphics[width=7cm]{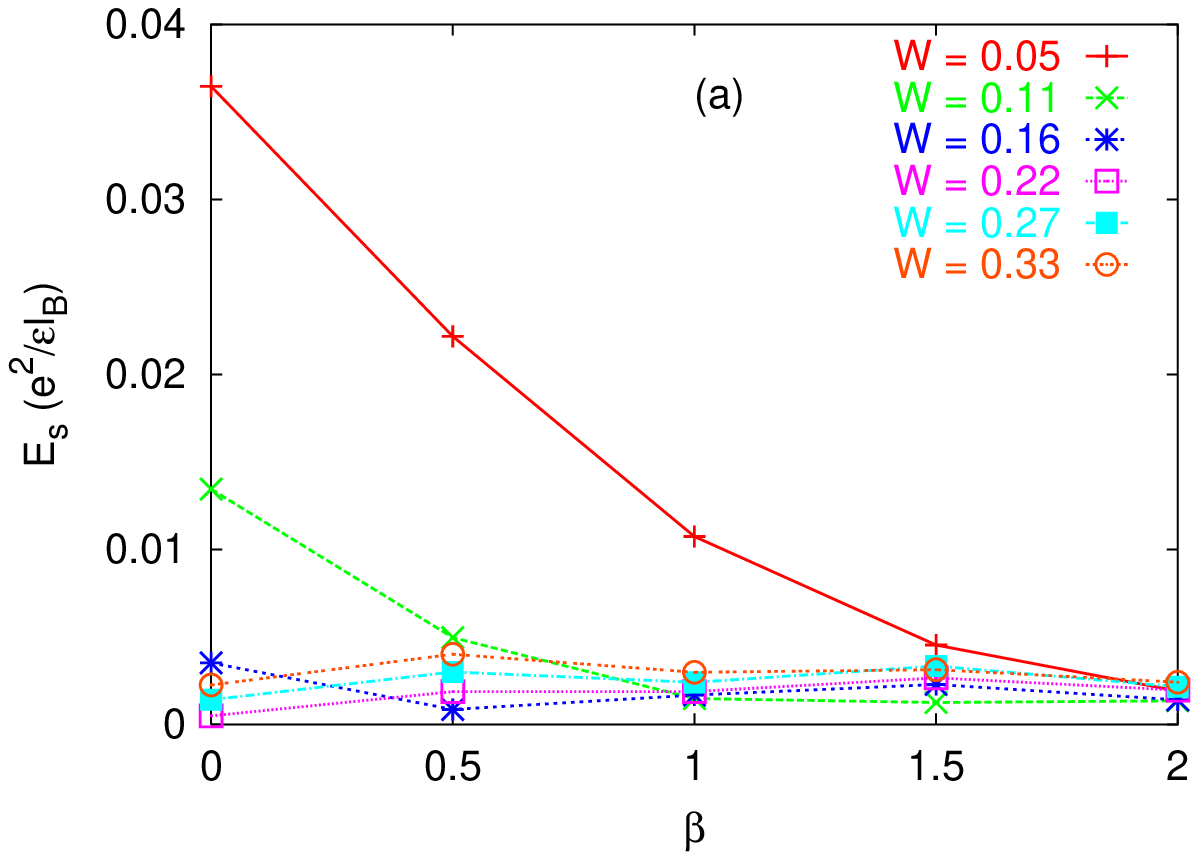} }
{\centering \includegraphics[width=7cm]{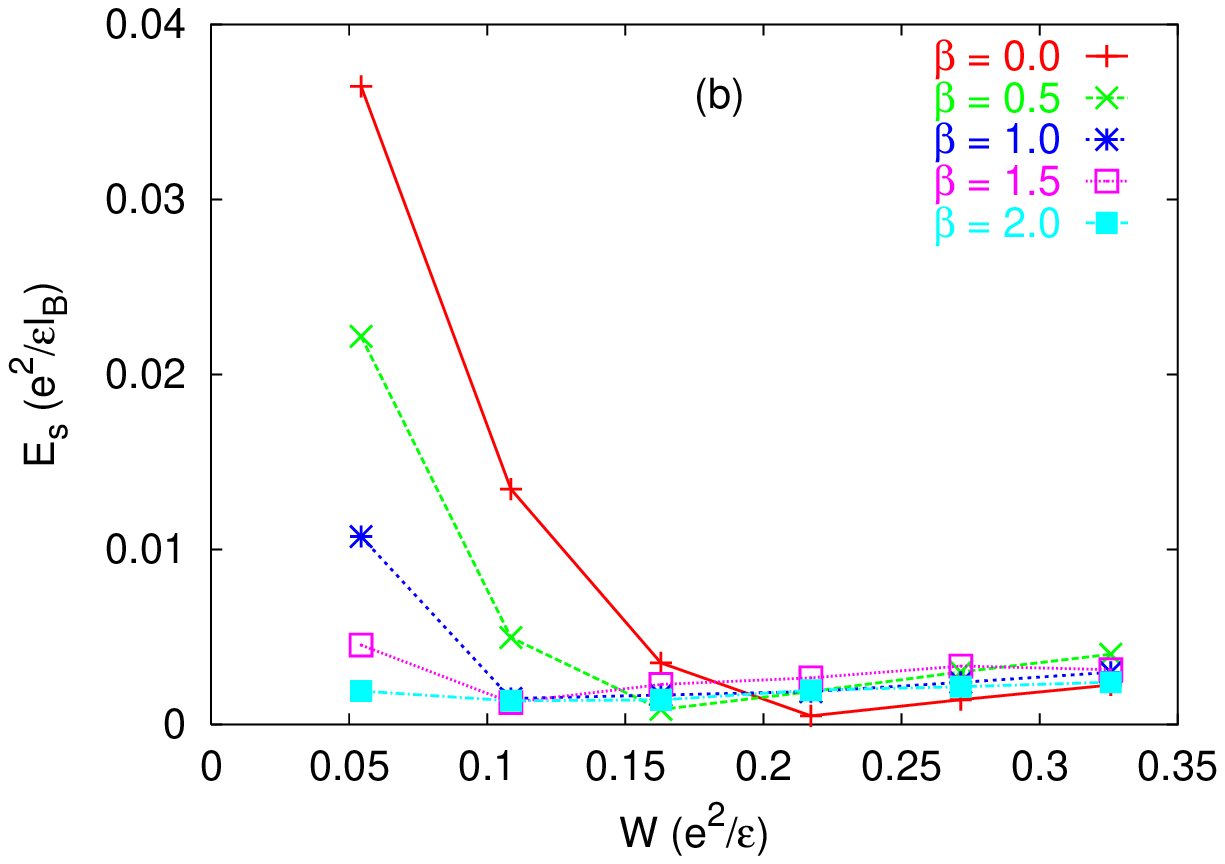} }
\caption{
\label{fig:spg}
Spectral gap, extrapolated to the thermodynamic limit, 
(a) as a function of $\beta$ for various $W$, and  
(b) as a function of $W$ for various $\beta$ at $\nu = 1/3$.  
}
\end{figure}

We extrapolate $E_s$ to the thermodynamic limit by fitting $E_s$ to 
\begin{equation}
E_s(N_e) = E_s(N_e \rightarrow \infty) + a_s / N_e.
\end{equation}
Figure~\ref{fig:spg}a shows $E_s(N_e \rightarrow \infty)$ as a function 
of $\beta$ for various $W$ at $\nu = 1/3$. 
For small $W$, $E_s$ decreases as $\beta$ increases, or as the
electron layer becomes thicker.
For large $W$, $E_s$ remains smaller than or close to 0.003, 
reflecting the closure of the spectral gap of the quantum Hall liquid.  
This finite but small residual value comes from the energy level spacing
of the insulating phase, which may disappear with a more sophisticate
definition of the spectral gap. 
In fact, for fixed layer thickness, $E_s$ first decreases with
increasing disorder, then increases weakly with disorder after
reaching its minimum. This minimum signals the critical disorder at
which the spectral gap closes in the thermodynamic limit.

\subsection{Groups of Chern numbers and their statistics}

Due to the topological three-fold degeneracy of the each energy level in
the thermodynamic limit for $\nu = 1/3$ quantum Hall liquids on a torus,
we define the $N_g$-th group of states as the $(3N_g-2)$-th,
$(3N_g-1)$-th, and $(3N_g)$-th states, and calculate the Chern number of
such a group $C(N_g)$ as the sum of the Chern numbers of the three
states within the group. 
While for small enough disorder, 
the states within each group are degenerate 
in the thermodynamic limit, this no longer holds 
for disorder strength large enough to distroy the
fractional quantum Hall phase and the accompanying topological order.

In earliar work,~\cite{sheng03} 
we found the following properties for the statistics of
the group Chern numbers for two-dimensional electrons at $\nu = 1/3$ 
with zero layer thickness. 

\begin{enumerate}

\item For weak disorder, the Chern number of the lowest group is always
unity, carried by the three lowest states. This together with the
fact that the three states become degenerate in the thermodynamic
limit, as well as the fact that there is a finite spectral gap
separating the three states to the rest, is the manifestation of the
$\nu = 1/3$ fractional quantum Hall state on a torus. In the
thermodynamic limit, each degenerate ground state carries a Hall
conductance of $e^2/3h$. This also holds for the Chern numers of upper
groups for small enough disorder.

\item Large enough disorder destroys the quantization of the Chern
number of each group in individual samples
from upper groups down to the lowest one. For
certain disorder, the probability of the group Chern number being unity
decreases sharply around one group, the energy of which we can 
define as the mobility edge. We will discuss the procedure in greater 
detail later in Sec.~\ref{sec:mobility_gap}.

\item The Chern number calculation is robust for small systems with as
few as 5 electrons. The fluctuation of the group Chern number has little
size dependence, which, therefore, can be used as an robust indicator 
of the degree of delocalization.

\end{enumerate}

\begin{figure}
{\centering \includegraphics[width=7cm]{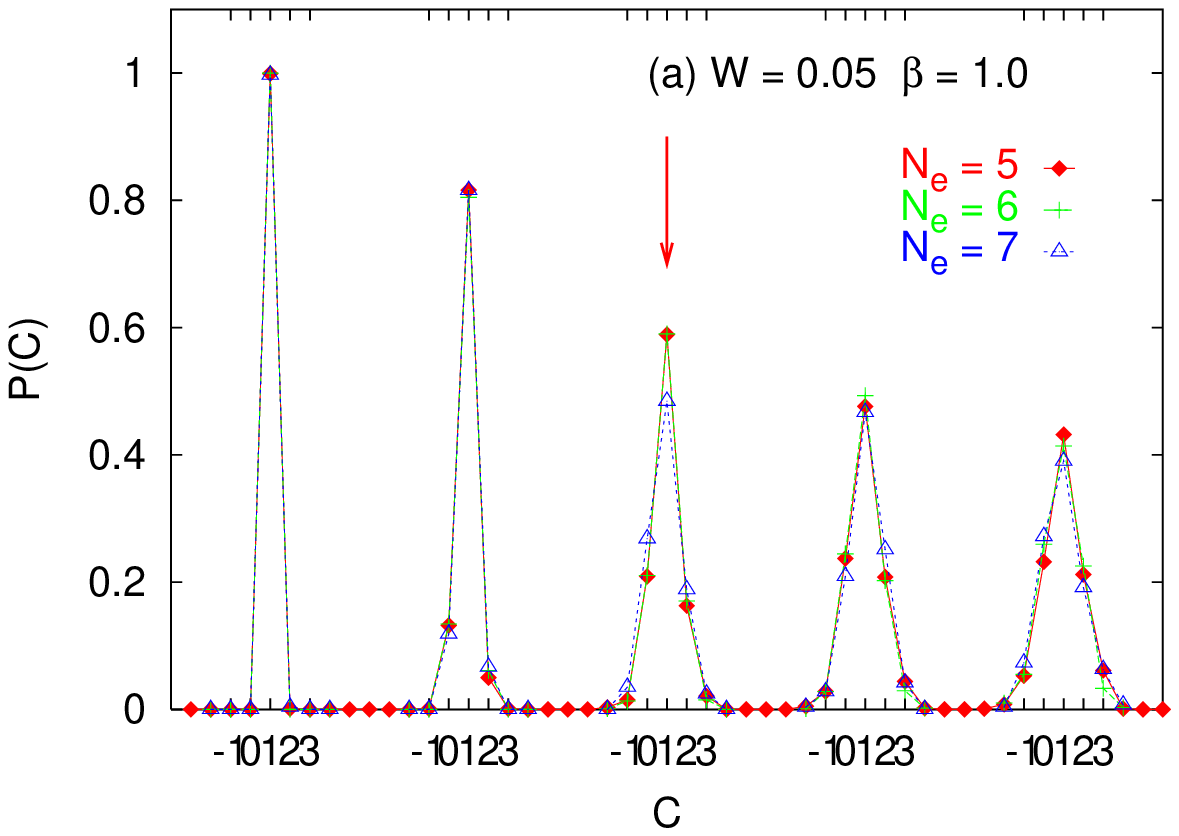} }
{\centering \includegraphics[width=7cm]{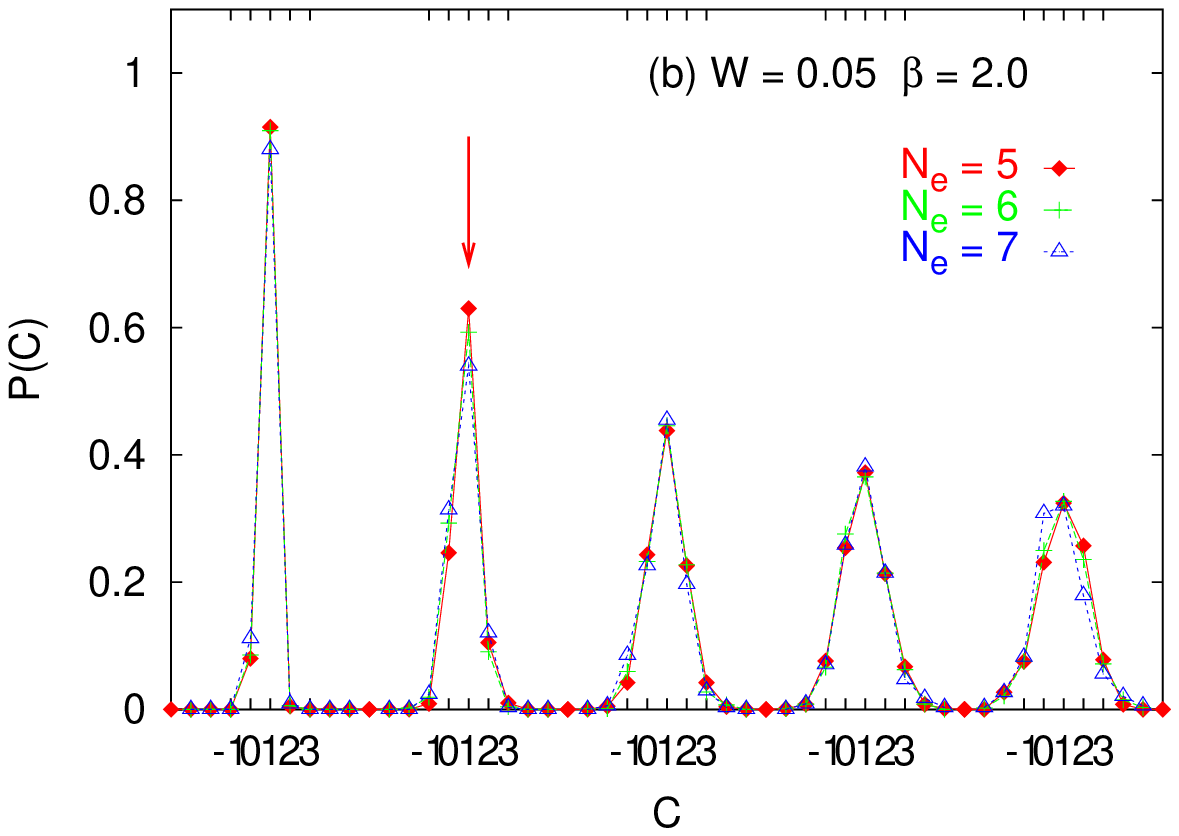} }
\caption{
\label{fig:u0.5_chn}
Probability distribution $P(C)$ of total Chern number $C$ for the lowest
five groups of states in systems of 5-7 electrons for $W = 0.05$ and for
$\beta$ = 1.0 and 2.0. Energy increases from left to right. The arrow in
each panel marks the group of states located at the mobility edge.}
\end{figure}

Figure~2 in Ref.~\onlinecite{sheng03} summarizes these properties, which
can be used to determine the mobility gap.  We repeat the calculation in
the presence of finite layer thickness.  Figure~\ref{fig:u0.5_chn}
shows, in analog to Fig.~2 in Ref.~\onlinecite{sheng03}, the probability
distribution $P(C)$ for group Chern number $C$ of the lowest 5 groups
of states in systems of 5-7 electrons for $W = 0.05$ and for
$\beta$ = 1.0 and 2.0.  We find that the above-mentioned
properties remain intact qualitatively in the presence of finite layer
thickness. Quantitatively, the increasing layer thickness shifts the
mobility edge toward the ground state. For example, for $\beta =
2.0$, the probility $P(C = 1)$ of the lowest group is close to unity, 
suggesting that fractional quantum Hall phase survives at $W = 0.05$.
Meanwhile, $P(C = 1)$ of the second lowest group drops sharply
to about 0.6, and may drop further lower for larger systems. This
indicates the location of the mobility edge.

\begin{figure}
{\centering \includegraphics[width=7cm]{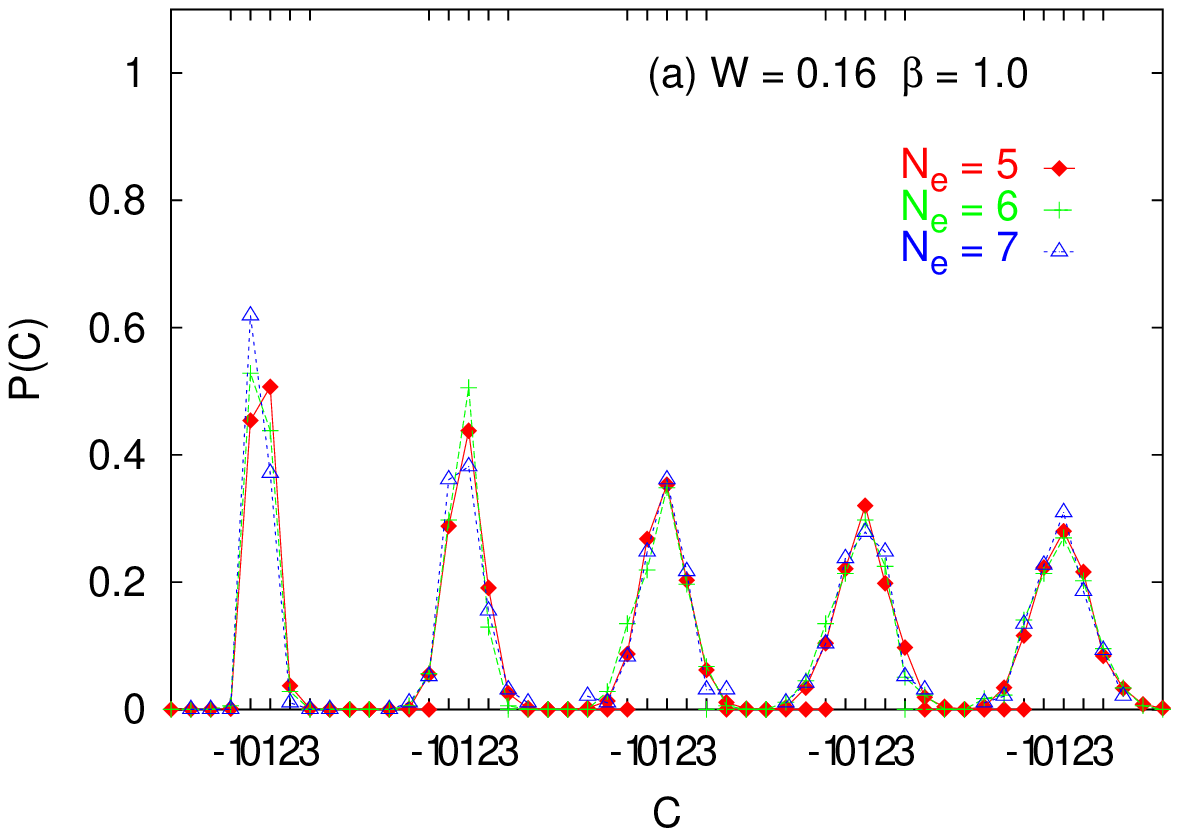} }
{\centering \includegraphics[width=7cm]{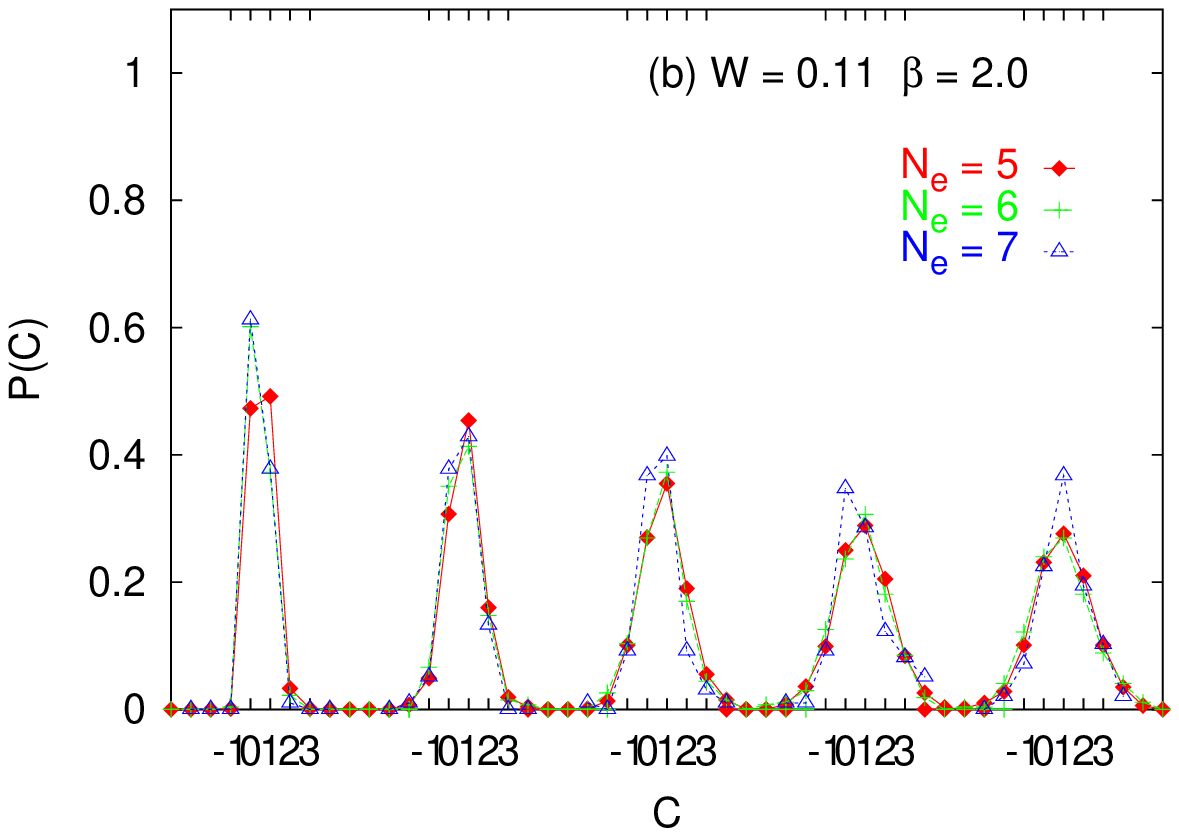} }
\caption{
\label{fig:ucrit_chn}
Probability distribution $P(C)$ of total Chern number $C$ for the lowest
five groups of states in systems of 5-7 electrons for
$W = 0.16$ and 0.11 for $\beta$ = 1.0 and 2.0, respectively. 
Energy increases from left to right. }
\end{figure}

We compare, in Fig.~\ref{fig:ucrit_chn}, the distribution $P(C)$ for
larger disorder $W = 0.16$ and 0.11 for $\beta$ = 1.0 and 2.0, 
respectively, again for the lowest 5 groups in 5- to 7-electron
systems. For the strong disorder, even the quantization of the lowest
group no longer holds, which becomes much smaller than 1,
indicating an insulating ground state in the thermodynamic limit. 

\subsection{Mobility gap}
\label{sec:mobility_gap}

As discussed above, the fluctuation of the total 
Chern numbers in each group of states is an indication of the degree of
delocalization, which we can used to determine the mobility edge.
Similar as in Ref.~\onlinecite{sheng03}, we define $P_{\rm ext} = 1 -
P(C = 1)$. The value of $P_{\rm ext}$ is the probability of the
breakdown of the Hall-conductance quantization and thus a measure of the
delocalization of the charge excitations. This is analogous to the
non-interacting IQHE case, where particle excitations from localized states
to delocalized states lead to fluctuation of the total Chern number, or
Hall conductance. 

\begin{figure}
{\centering \includegraphics[width=7cm]{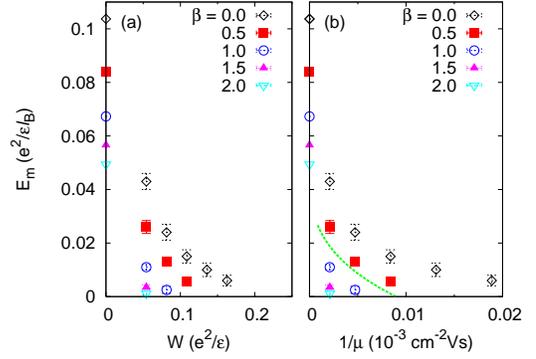} }
\caption{
\label{fig:mobility_gap}
(a) Mobility gap $E_m$ as a function of $W$ 
for various layer thickness $\beta$.
$E_m$ is extrapolated from systems of $N_e =$ 4-8 electrons 
to the limit $1/N_e \rightarrow 0$.
(b) Dependence of $E_m$ on inverse mobility $1/\mu^0$ for various $\beta$.
The dashed line is converted from a fit to experimental
data (taken from Ref.~\onlinecite{boebinger87}).
Here, we use an empirical mobility-density relation as well as
a mobility-disorder relation in the Born approximation
$\mu^0 = e \hbar^3 / (m^{*2} W^2)$. 
The data points on $y$-axes (for $W = 0$ and 1/$\mu = 0$)
are the gaps for pure systems, whose calculations 
are illustrated in detail in Appendix~\ref{sec:puregap}. } 
\end{figure}

For small disorder, $P_{\rm ext}$ remains close to 0 for groups of
states beyond the ground-state manifold. This reflects that the mobility
gap (which separates localized states from delocalized states) is
different from the spectral gap (which separates the ground-state
manifold to higher-energy states),  as excitations across spectral
gap may not lead to the fluctuations of the Hall conductance or
contribute to the longitudinal conductance.
Although the two gaps likely disappear
simultaneously when disorder is large enough to destroy the quantum Hall state. 
Therefore, we expect that $P_{\rm ext}$ rises sharply (probably
abruptly) at the mobility gap in the thermodynamic limit, a signature
imprinted in finite systems as well. Here, we find the sharpest jump of
$P_{\rm ext}$ between one group and its lower neighboring group, 
and define the energy of the higher group as the mobility edge. 
We measure the mobility gap $E_m(N_e)$ from the
ground-state energy to the mobility edge for the system of $N_e$
electrons. Since the states in each group
are degenerate in the thermodynamic limit, we use the average energy in
each group to calculate $E_m(N_e)$ to reduce finite-size fluctuations. We
then extropolate $E_m$ to the thermodynamic limit from $N_e = 4$-8
electrons. We plot the resulting $E_m$ as a function of $W$ for various
$\beta$ in Fig.~\ref{fig:mobility_gap}a. The
plot clearly demonstrates that finite layer thickness reduces the
mobility gap as expected. 
In order to show the overall trend in the disorder dependence of the gap,
particularly for weak disorder for various 2-D layer thicknesses, 
we include the gaps for pure systems on $y$-axes (for $W = 0$ and 1/$\mu
= 0$). 
The calculations of the pure gaps, which differ from the Chern number
calculations in disordered systems, are illustrated in detail in 
Appendix~\ref{sec:puregap}.
Comparing Fig.~\ref{fig:mobility_gap}a with Fig.~\ref{fig:spg}b, we find
that the mobility gap and the spectral gap differ significantly for small
disorder and small layer thickness, although they both decrease with
increasing disorder, as well as with increasing layer thickness.
The two gaps appear to disappear at roughly the same disorder strength.

In experiments, an energy gap $\Delta$ in the excitation spectrum 
of the correlated many-body ground state can be extracted from the
temperature dependence of the magnetoresistivity, $\rho_{xx} \propto
\exp (-\Delta / 2 k_B T)$, where $k_B$ is the Boltzmann's constant and 
$\Delta / 2$ is often referred as the activation 
energy~\cite{chang83,boebinger87,willett88}.
This activation energy is related to the mobility gap we calculated,
which separates the ground state from its delocalized quasiparticle
excitations.  
Boebinger {\it et al.}~\cite{boebinger87} systematically studied the
activation energy for $\nu = 1/3$, 2/3, 4/3, and 5/3 and its dependence
on sample mobility $\mu$ (an indication of disorder) in a series of
GaAs-Al$_x$Ga$_{1-x}$As samples. 
For a class of high-mobility (at that time - with mobility up to $10^6$
cm$^2$/V s) samples, they found that $\Delta \simeq 0.049 e^2/\epsilon l
- 6$ K, consistent with a simple phenomenological model~\cite{chang83} 
that assumes a disorder-broadened excitation energy level 
with half-width $\Gamma = 6$K.
Willett {\it et al.} studied a then untra-high-mobility sample and
compared the activation energy to theoretical results incorporating
finite layer thickness~\cite{zhang86} and Landau level
mixing.~\cite{yoshioka84,yoshioka86} 
While the agreement between experimental results and theoretical
caculations (in the absence of disorder) is satisfactory for magnetic
field stronger than 10 T, they diverge significantly at smaller magnetic
field, presumably due to disorder. 
Simple theories~\cite{macdonald85,gold86} with {\it ad hoc} treatment 
of disorder fail to account for all the discrepencies. 

With our numerical calculations which treat disorder and layer thickness
on an equal footing, we can now attempt to compare our results to
experimental ones quantitatively.
In experiments, the mobility $\mu$ dependence of $\Delta$ can then 
be extracted from the known dependence of $\mu$ on the electron
density $n$ of these samples, since $n$ determines the magnetic field
$B$ (or $l_B$) at the $1/3$ family of fillings. 
For semi-quantitative comparison, we use a typical
dependence, $\mu = \mu_0 (n / n_0)^{1.5}$, where $\mu_0 =$ 600,000
cm$^2$/V s and $n_0 = 1.5 \times 10^{11}$cm$^{-2}$, as extracted from
Fig.~1 of Ref.~\onlinecite{boebinger87}.
For comparison, we assume that in our simple disordered model both the
(zero field) mobility and the (high field) mobility gap are dominated by
short-range scatterers (appropriate for these then high-mobility
samples). 
In the Born approximation (as derived in Appendix~\ref{sec:mobility}), 
we have 
\begin{equation}
\mu^0 = e \hbar^3 / (m^{*2} W^2).
\end{equation}
Figure~\ref{fig:mobility_gap}b
compares this empirical formula of $\Delta(\mu^0)$ with the mobility gap
we obtained in our calculation for various $\beta$.
We find that the experimental data fall nicely into the range of 
$0.5 < \beta < 1$. 
This is fully expected in typical experiments: the variational parameter
$b^{-1}$ for a typical GaAs-AlGaAs sample with an electron density of
$N_0 = 10^{11}$ cm$^{-2}$ is close to the magnetic
length $l_B$,~\cite{zhang86} which scales with $B^{-1/2}$ 
(e.g. $l_B = 57$~\AA\ for $B = 20$~T).
In particular, Willett {\it et al.} obtained $b^{-1} = 39 \pm 1$~\AA\
for their sample, giving $\beta = (b l_B)^{-1} \approx 0.68$ 
at $B = 20$~T.
We do not include, however, the effects of Landau level mixing,  
which leads only to a small reduction of the mobility gap 
for clean samples.~\cite{yoshioka86,willett88} 

\section{Effects of Correlated Potential}
\label{sec:correlation}

In this section, we briefly discuss the effects of impurity potential with
finite correlation length for $\nu = 1/3$. 
The correlated potential is, in particular, 
relevant to ultra-high-mobility 2DEG, such as GaAs based systems 
in which impurities are introduced remotely above the 2DEG.
The mobility of these systems is generally believed to be limited 
by remote impurity scattering, 
rather than by short-range interface defects. 
To consider the correlated potential effects, we introduce the Gaussian
correlated random potential 
\begin{equation}
\langle U_{\bf q}U_{{\bf q}'}\rangle = 
{W^2 \over A} \delta_{{\bf q},-{\bf q'}} e^{-q^2 \xi^2 / 2},
\end{equation}
as described in detail in Sec.~\ref{sec:model}.
Here, we restrict ourselves, for simplicity, to bare Coulomb interaction
without considering finite layer thickness. 
We are mostly interested in the case of small correlation length $\xi
\sim 1$, in units of $l_B$, where quantitative results can be reached. 

\begin{figure}
{\centering \includegraphics[width=7cm]{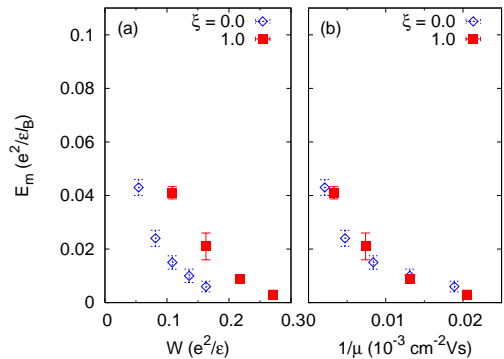} }
\caption{
\label{fig:mobility_gap_correlated}
(a) Mobility gap $E_m$ as a function of $W$ for Gaussian correlated
potential with $\xi = 1.0$, in units of $l_B$, 
compared with Gaussian white-noise potential ($\xi = 0$).
(b) Dependence of $E_m$ on inverse mobility $1/\mu$ for $\xi = 1.0$ and
0.0. 
}
\end{figure}

The results we find, in particular for $\xi = 1$, 
are qualitatively similar to those with Gaussian
white-noise potential.~\cite{sheng03}
Figure~\ref{fig:mobility_gap_correlated}a compares the mobility gap
$E_m$ as a function of $W$ for Gaussian white-noise potential ($\xi =
0$) and Gaussian correlated potential with $\xi = 1$.
The trend that increasing disorder strength destroys the mobility gap is
generically the same for impurity potential with or without
correlation. 
However, $E_m$ survives larger $W$ (by a factor of roughly 50\%) for $\xi
= 1$ than for $\xi = 0$. 
This is not surprising since impurity potential correlation enhances the
electron mobility, in particular for low electron densities. 
As shown in Appendix~\ref{sec:mobility}, the
mobility of the 2DEG, in the Born approximation, 
is enhanced by a factor of 
\begin{equation}
{\mu \over \mu^0} = 
{e^{k_F^2 \xi^2} \over I_0 (k_F^2 \xi^2) - L_0 (k_F^2 \xi^2)},
\end{equation} 
where $I_0(x)$ is the zeroth-order modified Bessel function of the first
kind and $L_0 (x)$ the zeroth-order modified Struve
function.
Here, $k_F = \sqrt{4 \pi n} = \sqrt{2 \nu} / l_B$ is the Fermi vector of
the polarized 2DEG. 
For $\xi = 1$, the factor is $\mu / \mu^0 = 2.556$. 
In the long-range limit, one obtains
\begin{equation}
{\mu \over \mu^0} \approx  \sqrt{8 \pi} (k_F \xi)^3.
\end{equation}

Figure~\ref{fig:mobility_gap_correlated}b shows the mobility gap $E_m$
as a function of $1 / \mu$ for impurity potentials with and without
correlation, as we have done here and in an earlier
paper.~\cite{sheng03}  
Interestingly, $E_m$ for $\xi = 1$ scales back and lies roughly on top of
the data for $\xi = 0$. 
This suggests that as long as $\xi$ is not too large, 
the effect of the range of potential can be lumped into that of 
sample mobility;
this makes comparisons between samples of
different type, as well as between theory and experiment more meaningful,
because the mobility is the directly measured character of a sample, while 
the details of disorder potential in general vary from sample to sample. 

We do not have, however, quantitative conclusion for vary large $\xi$,
which is believed to be responsible for those ultra-high-mobility
samples. 
Our finite-size calculations prevent us from considering large $\xi$
close to or even larger than the system size, at which stage the system
properties saturate.
We would also need to calculate Chern numbers for a lot more eigenstates
to obtain the mobility gap for the smooth potential, 
which is beyond our current computing capabilities.  
Nonetheless, it is expected that for those ultra-high-mobility
samples, disorder plays a less important role, and thus the mobility
gap depends less on the mobility of a sample.~\cite{willett88}

\section{Conclusions and discussions}
\label{sec:summary}

Semiconductor heterojunctions with modulation doping is the de facto
fabrication technique for high-mobility quasi-two-dimensional samples
commonly used for the study of the fractional quantum Hall effects.  The range
and strength of the impurity potential and the finite layer thickness of
the systems can significantly affect the transport properties
of the system, such
as the activation gap of the fractional quantum Hall liquids.  
In this paper, we have studied these issues using
a microscopic disordered model of the fractional quantum Hall liquids in
a toroidal geometry.
With the help of the Chern number calculation, 
which is capable of directly probing the
localization properties of many-body states, 
we have determined the mobility edge of the Hall liquids 
based on the fluctuations of the Hall conductivity 
and studied the dependence of mobility gap on
disorder strength, layer thickness, and potential correlation. 

Finite 2-D layer thickness has significant effects on the properties of
the 2DEG in experimental samples. The spread of the electron wave
function in the perpendicular direction softens the Coulomb interaction
between electrons, reflected both in the density-of-states and in the
ground state energy of the system. Consequently, the spectral gap, which
separates ground state manifold from excited states, decreases with
increasing layer thickness, as well as with increasing disorder
strength. The mobility gap, associated with the fluctuation of
topological Chern numbers - signaling delocalized excitations,
also decreases with increasing layer thickness and disorder. However,
the two gaps are different by definition and, indeed, distinguishable in
numerical calculations, in particular for small layer thickness and weak
disorder. Putting reasonable experimental parameters, we found our
results of the mobility gap are in excellent agreement with the
activation gap measured by experiments, suggesting that the mobility gap
is responsible for the activated behavior in longitudinal resistivity. 
Disorder and 2-D layer thickness are the two dominant factors affecting
the value of the gap. 

We also investigate the effect of the correlated disordered potential on the
mobility gap. For small correlation length, we found an enhancement in
the mobility gap compared to the case of uncorrelated potential for the
same disorder strength. Such enhancement is consistent with the
enhancement in mobility of the 2DEG in the presence of the correlated
potential. As long as the correlation length is small, the effect of the
correlated potential can be attributed to that of the sample mobility,
which demonstrates it meaningful to compare the gap as a function of
mobility among samples of different type and between theory and
experiment. 

In this paper, we model the electron wave function
in the perpendicular direction by the variational Fang-Howard wave
function. This is appropriate for modulation-doped
GaAs-Al$_{x}$Ga$_{1-x}$As heterojunctions, in which the Fang-Howard
function is a very good approximation to the numerical self-consistent
ground state wave function.~\cite{stern84} 
For realistic values of layer thickness, the qualitative nature of
the ground states and the low-energy excitations of the
quasi-two-dimensional systems remains
unchanged, in particular in the incompressible phases. 

Meanwhile, Shayegan {\it et al.}~\cite{shayegan90} studied the
fractional quantum Hall effects in thick parabolic quantum wells. 
Due to the selective doping of Al and screening, 
electrons experience a flat potential and, therefore, 
the electron density are roughly uniform in the quantum well. 
This results in a significantly larger layer thickness,
which increases with electron areal density. In such a system, Shayegan
{\it et al.}~\cite{shayegan90} observed a dramatic decrease in the
activation gap with increasing layer thickness and thus the collapse of
the fractional quantum Hall effect.  
He {\it et al.}~\cite{he90} 
studied the effects of layer thickness using a phenomenological model
potential (neglecting the accompanying effects of Landau level
mixing) and obtained qualitatively consistent results to the
experimental measurements. 
Although we have not repeated our calculations with a different wave
function more suitable for a parabolic quantum well, we expect a
similar trend of decreasing mobility gap with increasing layer thickness. 
We would like to point out that
Fig.~\ref{fig:mobility_gap} clearly demonstrates that increasing 
layer thickness can trigger a transition from a $\nu = 1/3$ 
fractional quantum Hall liquid to an insulator for fixed disorder strength. 
The transition becomes easier to occur in the presence of larger disorder. 

\acknowledgments

This work is supported by the Schwerpunktprogramme
``Quanten-Hall-Systeme'' der DFG (X.W.), 
ACS-PRF 41752-AC10 (D.N.S.), 
US DOE under contract DE-FG03-02ER-45981 (E.H.R.),
NSF grant DMR-0307170 (D.N.S.), DMR-0225698 (K.Y.),
and NSF under MRSEC grant DMR-0213706 at the Princeton
Center for Complex Materials (R.N.B. and F.D.M.H.). 
F.D.M.H. wishes to thank the KITP for support (NSF PHY99-07949) 
during his visit.

\appendix

\section{Ground state energy of a two-dimensional square Wigner lattice
with finite layer thickness}
\label{sec:squarelattice}

We calculate the ground-state energy of finite-size systems with
periodic boundary conditions. To compare our results with earlier
works~\cite{macdonald84,chakraborty86}, we must add, to our numerical
ground-state energy, a single-electron contribution from the interaction of
an electron and its images due to periodic boundary conditions
(see also Appendix~\ref{sec:puregap}).
In an ideal two-dimensional square system of linear size 
$L = \sqrt{2 \pi N_s} l_B$, 
this is simply the Madelung energy of a square lattice with
lattice constant $L$ and, for each electron, is~\cite{yoshioka83,su84} 
\begin{eqnarray}
\label{eqn:bonsall}
\epsilon_M 
&=& -{e^2 \over L} \left ( 2 - \sum_{l_1,l_2} \!^{'} \phi_{-1/2} 
[\pi (l_1^2 + l_2^2)] \right ) \nonumber \\
&=& - {3.9 e^2 \over 2L}
\end{eqnarray}
calculated first in the context of the Coulomb energy of
the two-dimensional classical Wigner crystal.~\cite{bonsall77}
The summation is performed over lattice sites 
${\bf l} = l_1 {\bf a}_1 + l_2 {\bf a}_2$ for the primitive lattice
vectors ${\bf a}_1$ and ${\bf a}_2$ except ${\bf l} = 0$ 
and the $\phi_n(x)$ are the Misra functions
\begin{equation}
\phi_n(x) = \int_1^{\infty} dt \, t^n e^{-xt}.
\end{equation}  
The factor of $1/2$ in Eq.~(\ref{eqn:bonsall})
comes from the double counting of 
the electron-electron interaction. 

In the presence of finite layer thickness, 
the Coulomb energy of the quasi-two-dimensional classical
Wigner crystal is also known for generic lattices.~\cite{fujiki92}
The wave function $\phi(z)$ in the perpendicular direction 
enters through the following function 
\begin{equation}
f(y, b) = \int_0^{\infty} dz \int_0^{\infty} dz'
\phi^2(z) \phi^2(z') e^{-y |z-z'|^2}
\end{equation}
and for the Fang-Howard wave function 
\begin{equation}
\phi(z) = (b^3/2)^{1/2} z e^{-bz/2},
\end{equation}
one can obtain $f(y, b) = \tilde{f}(b^2/4y)$, where
\begin{equation}
\tilde{f}(t) =  
{3t \over 4} -{t^2 \over 2} + 
\sqrt{t} e^t 
\left ({3\over 8}-{t \over 2}+{t^2 \over 2} \right )
\Gamma \left ({1 \over 2}, t \right ) .
\end{equation}
Here, the incomplete gamma function $\Gamma (a,x)$ is given by
\begin{equation}
\Gamma(a,x) \equiv \int_x^{\infty} dt\, t^{a-1}e^{-t}
=x^a \phi_{a-1} (x).
\end{equation}
For $a = 1/2$, the incomplete gamma function is also 
related to the complementary error function
\begin{equation}
{\rm erfc}(z) = {2 \over \sqrt{\pi}} 
\int_z^{\infty} dt\, e^{-t^2} 
= {\Gamma \left ({1 \over 2}, z^2 \right) \over \sqrt{\pi} },
\end{equation}
for $z > 0$. 

With the help of an integral transform, one can replace
the slowly converging sum of Coulomb energy into two rapidly converging
sums: one for the short-range part and the other the long-range, in the
same spirit as in the original Ewald method.~\cite{bonsall77} 
In the following, the parameter that separates the two sums is denoted
as $y_0$, chosen to be $\pi / L^2$ in our calculation.  
For convenience, one introduces the Jacobi $\theta$ function 
\begin{equation}
\theta(z, X)= \sum_{m = -\infty}^{\infty} 
e^{2 \pi m z} e^{-\pi m^2 X},
\end{equation}
which converges fast for not too small $X$. 
Here, we only review the results for a square lattice with lattice
constant $L$ for simplicity. 
To be specific, they are the $q \rightarrow 0$ limit of
Eqs.~(16)-(21) in Ref.~\onlinecite{fujiki92} for the square lattice.
The Coulomb energy per electron for the square lattice 
can be written as 
\begin{equation}
N_e^{-1} E_{ee} = N_e^{-1} E_{ee}^> + N_e^{-1} E_{ee}^<,
\end{equation}
where
\begin{equation}
{E_{ee}^> \over N_e} = {e^2 \over 2 \sqrt{\pi}} 
\int_{y_0}^{\infty} dy \, y^{-1/2} f(y, b) \left [ 
\theta^2 \left ( 0; {L^2 y \over \pi} \right ) - 1 \right ]
\end{equation}
and 
\begin{eqnarray}
{E_{ee}^< \over N_e} &=& {\sqrt{\pi} e^2 \over 2 L^2} 
\int_0^{y_0} dy \, y^{-3/2} f(y, b) \left [ 
\theta^2 \left ( 0; {\pi \over L^2 y} \right ) - 1 \right ] 
\nonumber \\
&&- {\sqrt{\pi} e^2 \over 2 L^2} \int_{y_0}^{\infty} dy \,
y^{-3/2} f(y, b) \nonumber \\
&&-{e^2 \over 2 \sqrt{\pi}} \int_0^{y_0} dy \,  
y^{-1/2} f(y, b) \nonumber \\
&&+ N_e^{-1} E_{ee}^{hom}(q)|_{q=0},
\end{eqnarray}
where $E_{ee}^{hom}(q)$ is the Coulomb energy corresponding to a
homogeneous distribution
\begin{eqnarray}
N_e^{-1} E_{ee}^{hom}(q) &=& 
{\sqrt{\pi} e^2 \over 2 L^2} \int_0^{y_0} dy \, 
y^{-3/2} f(y, b) e^{-q^2/4y} \nonumber \\
&=& {\pi e^2 \over L^2} \left (
\left . {1 \over q} \right |_{q=0} - {15 \over 8b} \right ).
\end{eqnarray}
Here the singular $1/q$ term has its origin in the lack of charge
neutrality considered here. In fact, only nonsingular terms survive 
in the total
Coulomb energy once neutralizing background charge 
(e.g., located uniformly at $z = -b_d$) is present. 
The second term, as well as additional nonsingular 
terms introduced by the
interactions between electrons and the background charge, 
depends only on ``external'' parameters, such as
layer thickness $b$ and location of the background charge $b_d$. 
We neglected these terms since they have no effects 
on the results of the finite-size scaling 
in the $1/N_e \rightarrow 0$ limit. 

\section{Calculations of the Gaps in Pure Systems with Layer Thickness}
\label{sec:puregap}

In order to show the overall trend in the disorder dependence of the gap,
particularly for weak disorder for various 2-D layer thicknesses, 
we presented the pure gaps ($1/\mu=0$, $W=0$) in
Fig.~\ref{fig:mobility_gap} for $\nu = 1/3$.  
In this appendix we give the main details.
These types of calculations in PBC geometry have so far been largely
avoided for fear of strong finite size effects. 
These result from the interaction of the quasiparticles with
their images.  Indeed such effects remain
substantial (as large as 30\% in some cases) for even the largest sizes 
in exact diagonalization studies.  In order to remove these we
follow the practice of not neutralizing the quasiparticle (qp) and 
quasihole (qh) excitations\cite{haldane85}.  
This gives a positive contribution of
\begin{equation}
{e^{*2}\over 2\epsilon A}\int d^2\vec{r}V(r),
\end{equation}
where $e^*=e/3$ is the charge of the quasiparticle excitation, 
$A$ is the area of the system, and
\begin{equation}
V(r)=\int d^2\vec{q} {2\pi \over q}\exp\{i\vec{q} \cdot \vec{r}\},
\end{equation}
where $\vec{q}=n_1\vec{G}_1+n_2\vec{G}_2$ is the wave vector appropriate
for the PBC unit cell, $\vec{G}$'s are the corresponding reciprocal
lattice vectors, and $n$'s are integers. We next subtract the repulsive
interaction energy of a single quasiparticle with its images:
\begin{equation}
{e^{*2}\over 2 \epsilon}\sum_{\ell_1,\ell_2=-\infty}^{\infty}  
{1\over |\vec{X}(\vec{\ell})|},
\end{equation}
where $\vec{X}(\vec{\ell})=\ell_1 \vec{L}_1+\ell_2\vec{L}_2$, 
$\ell$'s are integers and $\vec{L}$'s are the direct lattice vectors.  
These two terms together add to 
\begin{equation}
\delta E= (e^*/e)^2 |S|=|S|/9,
\end{equation}
where $S$ is the classical ground state energy per electron 
(Madelung energy) of a Wigner crystal of 
electrons\cite{bonsall77,yoshioka84b}.  
Note that we are treating the quasiparticles as point objects.  
This approximation only introduces another finite size effect 
since the size of the unit cell is several magnetic lengths whereas the
substantial density variation of the quasi-particle excitations 
occurs over a magnetic length.

Starting from the usual expression of the gap in terms of 
the ground state energies:
\begin{eqnarray}
\Delta &=& E(\nu=1/3+qp) + E(\nu=1/3+qh) \nonumber \\
&& - 2E(\nu=1/3). 
\end{eqnarray}
With the above subtractions we obtain:
\begin{eqnarray}
\Delta^* &=& E^*(\nu=1/3+qp)+E^*(\nu=1/3+qh) \nonumber \\
&&-2E^*(\nu=1/3)+\delta E_{qp}+ \delta E_{qh},
\end{eqnarray}
where $E^*$ is the finite part of the energy without the Madelung term
($E^*=E-NS$) (note $S$ is negative).
This contribution taken together for the 3 ground state energies in
$\Delta^*$ is a finite size effect as are both $\delta E$ 
energies so they will disappear in the
thermodynamic limit ($\Delta^*_\infty=\Delta_\infty$). 
Therefore we need not correct 
any of these energies for finite layer thickness. 
We then extrapolate $\Delta^*$ to the thermodynamic limit 
from its $1/N$ dependence. 
We used the hexagonal unit cell for this part as it has 
the highest degree of symmetry.  
Table 1 gives the numerical values of the $\Delta_\infty$ 
that we have obtained from $N=7-10$ size systems.   
In the same table we give the corresponding results for the finite
layer systems.  We have shown all these on the y-axis of 
Fig.~\ref{fig:mobility_gap} (for 1/$\mu = 0$ and $W=0$).  

\begin{table}
\label{tab:pure_gap}
\begin{tabular}{cccccc}
\hline\hline
$\beta$  \hspace{0.2cm} & 0.0 & 0.5 & 1.0 & 1.5 & 2.0 \\
\hline
$\Delta$ \hspace{0.2cm} & 0.1037 \hspace{0.2cm} & 0.08410 \hspace{0.2cm} 
& 0.06729 \hspace{0.2cm} & 0.05656 \hspace{0.2cm} & 0.04940 \\
\hline\hline
\end{tabular}
\caption{Gaps in pure FQHE systems at $\nu = 1/3$ 
for various 2-D layer thickness.}
\end{table}

\section{Mobility of two-dimensional non-interacting electrons in random
potential} 
\label{sec:mobility}

In this appendix, we review the results of the mobility of a
two-dimensional non-interacting electron system with a random potential
in standard perturbation theory~\cite{mirlin96} and apply them to our
model system. 

Consider spinless electrons in a quenched random potential $U(r)$ with
Gaussian correlation
\begin{equation}
\langle U({\bf r}) U({\bf r}') \rangle 
= {W^2 \over 2 \pi \xi^2} e^{- |{\bf r} - {\bf r}'|^2 / 2 \xi^2} 
\equiv W(|{\bf r} - {\bf r}'|),
\end{equation}
where $\xi$ is the characteristic correlation length. Note that in the
limit of $\xi \rightarrow 0$, we recover 
\begin{equation}
\langle U({\bf r}) U({\bf r}') \rangle = W^2 \delta ({\bf r} - {\bf r}')
\end{equation}
for a Gaussian white-noise potential. 
The total scattering rate in the Born approximation is~\cite{mirlin96} 
\begin{eqnarray}
\label{eqn:scattering}
{1 \over \tau_{tr}} 
&=& {2\pi \rho \over \hbar} \int_0^{2\pi} {d\phi\over 2\pi}
\tilde{W}(2k_F\sin{\phi\over 2})(1-\cos\phi) \\
&=& {4 \pi^2 \rho \over \hbar} 
\int_0^\infty dr\, r [J_0^2(k_Fr)-J_1^2(k_Fr)]W(r), 
\nonumber
\end{eqnarray}
where $\rho = m^* / (2 \pi \hbar^2)$ is the density of states of free
electrons and $k_F = \sqrt{4 \pi n} = \sqrt{2 \nu} / l_B$ 
is the Fermi vector. $\tilde{W}(k)$ is the Fourier
transformation of $W(r)$
\begin{equation}
\tilde{W}(k) = \int d^2 r W(r) 
\exp (-i{\bf k} \cdot {\bf r}).
\end{equation}
For the Gaussian white noise potential, $\tilde{W}(k) = W^2$, 
thus
\begin{equation}
{1 \over \tau_{tr}} = {m^* W^2 \over \hbar^3}.
\end{equation}
The mobility of the system with the short-range potential is, therefore,
\begin{equation}
\mu^0 = {e \tau_{tr} \over m^*} = {e \hbar^3 \over m^{*2} W^2 }.
\end{equation}
With simple algebras, we can rewrite the mobility $\mu$ in terms of
the cyclotron energy $\hbar \omega_c$, 
the Coulomb energy $e^2/\epsilon l_B$, 
and the magnetic length $l_B$ as
\begin{equation}
\mu^0 = {e \over \hbar} l_B^2 
\left [ {\hbar \omega_c \over e^2/\epsilon l_B}
\right ]^2 {1 \over \bar{W}^2},
\end{equation}
where $\bar{W} = \epsilon W / e^2$ is the dimensionless disorder
strength.

For a long-range potential, $k_F \xi \gg 1$, one finds~\cite{mirlin96}
\begin{equation}
{1 \over \tau_{tr}} = -{m \over (\hbar k_F)^3}
\int_0^\infty dr {W'(r)\over r}.
\end{equation}
In particular, for the Gaussian correlated potential,
\begin{equation}
\tilde{W}(k) = W^2 e^{-k^2 \xi^2 / 2},
\end{equation}
we can integrate Eq.~(\ref{eqn:scattering}) and obtain 
\begin{equation}
{1 \over \tau_{tr}} = {m^* W^2 \over \hbar^3} 
e^{-k_F^2 \xi^2} \left[I_0 (k_F^2 \xi^2) - L_0 (k_F^2 \xi^2) \right ],
\end{equation}
where $I_0(x)$ is the zeroth-order modified Bessel function of the first
kind and $L_0 (x)$ the zeroth-order modified Struve
function.~\cite{abramowitzbook}
The mobility of the system with the Gaussian correlated potential is,
therefore,
\begin{equation}
\mu = \mu^0 
{e^{k_F^2 \xi^2} \over I_0 (k_F^2 \xi^2) - L_0 (k_F^2 \xi^2)},
\end{equation}
In the long-range limit, we obtain
\begin{equation}
{\mu \over \mu^0} \simeq  \sqrt{8 \pi} (k_F \xi)^3.
\end{equation}


\begin{thebibliography}{99}
\bibitem{fujiki92}
N.~M. Fujiki and D.~J.~W. Geldart,
Phys. Rev. B {\bf 46}, 9634 (1992).
Note that the definition of the layer thickness 
parameter $b$ is different from the one used in this work.
\end{thebibliography}
\end{document}